\documentclass[aps,prd,twocolumn,floatfix,noshowpacs,tightenlines,noshowkeys,superscriptaddress,amsmath,amssymb,nofootinbib]{revtex4}
\usepackage{amssymb,amsbsy,epsfig,color,graphicx}
\usepackage{color}
\usepackage{[longtable}
\usepackage{array}
\usepackage{dcolumn}   
\usepackage{cellspace}
\usepackage{mathtools}
\usepackage{amstext}
\usepackage{amssymb}
\usepackage{stmaryrd}
\usepackage{stackrel}
\usepackage{graphicx}
\usepackage{esint}
\usepackage[utf8]{inputenc}
\usepackage{blindtext}
\usepackage{float}
\restylefloat{table}
\usepackage{booktabs}
\usepackage{enumitem} 

\usepackage{etoolbox} 
\usepackage{lipsum} 
\usepackage[capitalize]{cleveref}
\usepackage{multirow}
\usepackage[caption=false]{subfig}
\renewcommand\[{\begin{equation}}
\renewcommand\]{\end{equation}}

\newcommand{\ba}{\begin{eqnarray}}
\newcommand{\ea}{\end{eqnarray}}

\makeatletter

\appto{\appendix}{%
\@ifstar{\def\theequation@prefix{A.}}%
{}%
}
\makeatother

\begin{document}

\title{Casimir effect in quadratic theories of gravity}

\author{Luca Buoninfante}
\affiliation{Dipartimento di Fisica "E.R. Caianiello", Universit\`a di Salerno, I-84084 Fisciano (SA), Italy}
\affiliation{INFN - Sezione di Napoli, Gruppo collegato di Salerno, I-84084 Fisciano (SA), Italy}
\affiliation{Van Swinderen Institute, University of Groningen, 9747 AG, Groningen, The Netherlands}

\author{Gaetano Lambiase}
\affiliation{Dipartimento di Fisica "E.R. Caianiello", Universit\`a di Salerno, I-84084 Fisciano (SA), Italy}
\affiliation{INFN - Sezione di Napoli, Gruppo collegato di Salerno, I-84084 Fisciano (SA), Italy}

\author{Luciano Petruzziello}
\affiliation{Dipartimento di Fisica "E.R. Caianiello", Universit\`a di Salerno, I-84084 Fisciano (SA), Italy}
\affiliation{INFN - Sezione di Napoli, Gruppo collegato di Salerno, I-84084 Fisciano (SA), Italy}

\author{Antonio Stabile}
\affiliation{Dipartimento di Fisica "E.R. Caianiello", Universit\`a di Salerno, I-84084 Fisciano (SA), Italy}
\affiliation{INFN - Sezione di Napoli, Gruppo collegato di Salerno, I-84084 Fisciano (SA), Italy}


\begin{abstract}
In this paper, we study the Casimir effect in a curved spacetime described by gravitational actions quadratic in the curvature. In particular, we  consider the dynamics of a massless scalar field confined between two nearby plates and compute the corresponding mean vacuum energy density and pressure in the framework of quadratic theories of gravity. Since we are interested in the weak-field limit, as far as the gravitational sector is concerned we work in the linear regime. Remarkably, corrections to the flat spacetime result due to extended models of gravity (although very small) may appear at the first-order of our perturbative analysis, whereas general relativity contributions start appearing at the second order. Future experiments on the Casimir effect might represent a useful tool to test and constrain extended theories of gravity.
\end{abstract}

\maketitle


\section{Introduction}\label{intro}

Einstein's general relativity (GR) has undergone many challenges in the last century, but it always proved its worth thanks to high precision experiments which have confirmed a plethora of its predictions~\cite{-C.-M.}. Despite its extraordinary achievements, there are still open questions which need to find an answer, that is nowhere to be found in GR. For instance, in the short-distance (ultraviolet) regime, from a classical point of view Einstein's theory turns out to be incomplete due to the presence of black holes and cosmological singularities, whereas from a quantum perspective it fails to be a renormalizable theory. On the other hand, at large scales GR is not capable of coming up with an explanation for dark matter and dark energy, even though their presence is strongly supported by observational data already available for a long time. 

In the past years, all these fundamental issues stimulated a vivid investigation revolving around a plausible extension of Einstein's GR domain. Among all the theories that popped out with the above intent, one of the straightforward approach consists of generalizing the Einstein-Hilbert action by including terms which are quadratic in the curvature, for example $\mathcal{R}^2,$ $\mathcal{R}_{\mu\nu}\mathcal{R}^{\mu\nu}$ and $\mathcal{R}_{\mu\nu\rho\sigma}\mathcal{R}^{\mu\nu\rho\sigma}.$ First interesting results in the context of quadratic gravity can be attributed to Stelle~\cite{-K.-S.}, who showed that a gravitational theory described by the Einstein-Hilbert action with the addition of $\mathcal{R}^2$ and $\mathcal{R}_{\mu\nu}\mathcal{R}^{\mu\nu}$ turns out to be power-counting renormalizable. However, this apparatus lacks of predictability due to the presence of a massive spin-$2$ ghost degree of freedom which breaks unitarity at the quantum level, when the standard quantization prescription is adopted \footnote{See Refs. \cite{Anselmi,Anselmi:2017ygm} for recent investigations where the theory is made unitary using an alternative quantization prescription where the massive spin-$2$ is not a ghost but becomes a {\it fake} degree of freedom ({\it fakeon}) which preserves the optical theorem.}. Despite the presence of the ghost field, such a theory can still be considered predictive as an effective field theory whose validity is accurate at energy scales lower than the cut-off represented by the mass of the ghost. Another important improvement of quadratic gravity can be found in the Starobinski-model of inflation~\cite{starobinski}, which is able to suitably explain the current data; differently from the model of Stelle, here only the term $\mathcal{R}^2$ shows up in the quadratic part of the action. It is also worthwhile to highlight that gravitational actions with quadratic curvature corrections were taken into account in several different frameworks (see for example Refs.~\cite{capoz1,lamb1,lamb2,radicella,capoz3,calchi,capoz4,capoz5,Asorey:1996hz,stabile_stabile_cap,stabstab,LamMohSta}).

The models so far discussed have been developed employing {\it local} quadratic theories of gravity, which means that the corresponding Lagrangian depends polynomially on the fields and their derivatives. In recent years, also {\it non-local} quadratic theories have aroused a significant interest, since the presence of non-local form-factors in the gravitational action may be useful both to solve the problem of ghosts and to considerably improve the ultraviolet behavior of the quantum theory (see Refs.~\cite{Tomboulis:1997gg,Biswas:2005qr,Modesto:2011kw,Biswas:2011ar,Biswas:2013cha,Biswas:2016etb,Koshelev:2017bxd} for more details).

On the other hand, all the models introduced up to now can be argument of further theoretical treatment. Indeed, as already anticipated, they can be included in countless applications of the most disparate physical frameworks. For what concerns our work, we are mainly focused on the analysis of the Casimir effect when the spacetime background is described by quadratic theories of gravity. The Casimir effect is a concrete manifestation of quantum field theory (QFT), which occurs whenever a quantum field is bounded in a finite region of space. Since it was firstly introduced to the scientific community~\cite{casimir}, it has risen a constant interest and investigative efforts, due to the possibility of extrapolating substantial pieces of information from experiments. Such a statement holds true not only for the case in which the Casimir effect is analyzed in flat spacetime~\cite{flat}, but also when the confinement of the quantum field occurs in a curved background~\cite{Sorge,curved}, and even when Lorentz symmetry is violated~\cite{lv,lv1} or mixed fields are considered \cite{mixed-fields}.

In this article, we study the Casimir effect in a curved spacetime emerging from a pure gravitational action quadratic in the curvature invariants. For this purpose, we closely follow the approach introduced in Ref.~\cite{Lambiase:2016bjy}, in which the authors analyze a scalar-tensor fourth order action stemmed from a non-commutative geometric theory. In contrast, we analyze the dynamics of a massless scalar field between two nearby plates in a curved background described by quadratic theories of gravity and in the weak-field limit. 

The paper is organized as follows: in Section II, we briefly review the most important features of gravitational theories whose action is quadratic in the curvature invariants, with a peculiar attention to the linearized solutions. In Section III, we study the dynamics of the massless scalar field in the context of the Casimir effect with a curved background. Section IV is devoted to the calculation of the main physical quantities of the Casimir effect, namely the mean vacuum energy density and the pressure, for several quadratic theories of gravity. Section V contains discussions and conclusion. 

Throughout the whole paper, the adopted metric signature is $\mathrm{diag}\left(-,+,+,+\right)$ and natural units $c=\hbar=1$ are used.

\section{Quadratic theories of gravity}

Let us consider the most general gravitational action which is quadratic in the curvature, parity-invariant and torsion-free \cite{Asorey:1996hz,Modesto:2011kw,Biswas:2011ar,Biswas:2013cha,Biswas:2016etb}
\begin{equation}
\!\!\!\!\begin{array}{rl}
S=& \displaystyle \frac{1}{2\kappa^2}\int d^4x\sqrt{-g}\left\lbrace \mathcal{R}+\frac{1}{2}\left(\mathcal{R}\mathcal{F}_1(\Box)\mathcal{R}+\mathcal{R}_{\mu\nu}\mathcal{F}_2(\Box)\mathcal{R}^{\mu\nu}\right.\right.\\[4mm]
& \displaystyle  \,\,\,\,\,\,\,\,\,\,\,\,\,\,\left.\left.+\mathcal{R}_{\mu\nu\rho\sigma}\mathcal{F}_3(\Box)\mathcal{R}^{\mu\nu\rho\sigma}\right)\right\rbrace,
\end{array}
\label{quad-action}
\end{equation}
where $\kappa:=\sqrt{8\pi G}$, $\Box=g^{\mu\nu}\nabla_{\mu}\nabla_{\nu}$ is the d'Alembert operator in curved spacetime and the form-factors $\mathcal{F}_i(\Box)$ are generic operators of $\Box$ that can be either local or non-local
\begin{equation}
\mathcal{F}_i(\Box)=\sum\limits_{n=0}^{N}f_{i,n}\Box^n,\,\,\,\,\,\,\,\,\,i=1,2,3.
\end{equation}
In what follows, we deal with both positive and negative powers of the d'Alembertian, which means we consider both ultraviolet and infrared modifications of Einstein's general relativity. Note that, if $n>0$ and $N$ is finite (namely, $N<\infty$), we have a local theory of gravity of order $2N+2$ in derivatives, whereas if $N=\infty$ and/or $n<0$ we have a non-local theory of gravity whose form-factors $\mathcal{F}_i(\Box)$ are {\it not} polynomials of $\Box.$

Our primary aim is to study the Casimir effect between two plates in a {\it slightly} curved background described by the action in Eq.~\eqref{quad-action}. Thus, we can apply a weak-field approximation in order to derive the linearized regime of Eq.~\eqref{quad-action} around the Minkowski background $\eta_{\mu\nu}$
\begin{equation}
g_{\mu\nu}=\eta_{\mu\nu}+\kappa h_{\mu\nu}\,,\label{lin-metric}
\end{equation}
where $h_{\mu\nu}$ is the linearized metric perturbation. 

At the linearized level, the relevant contribution coming from the action is of the order $\mathcal{O}(h^2)$; in such a regime, the term $\mathcal{R}_{\mu\nu\rho\sigma}\mathcal{F}_3(\Box)\mathcal{R}^{\mu\nu\rho\sigma}$ in Eq.~\eqref{quad-action} can be safely neglected. Indeed, if we do not exceed the aforementioned order of expansion, it is always possible to rewrite the Riemann squared contribution in terms of Ricci scalar and Ricci tensor squared by virtue of the following identity:
\begin{equation}
\mathcal{R}_{\mu\nu\rho\sigma}\Box^n\mathcal{R}^{\mu\nu\rho\sigma}=4\mathcal{R}_{\mu\nu}\Box^n\mathcal{R}^{\mu\nu}-\mathcal{R}\Box^n\mathcal{R}+\mathcal{O}(\mathcal{R}^3)+{\rm div},
\end{equation}
where {\rm div} stands for total derivatives and $\mathcal{O}(\mathcal{R}^3)$ only contributes at order $\mathcal{O}(h^3).$ Hence, in the linearized regime we can set $\mathcal{F}_3(\Box)=0$ without loss of generality.

\subsection{Linearized metric solutions: point-like source in a weak-field limit}

We now want to linearize the action Eq.~\eqref{quad-action} and analyze the corresponding linearized field equations. By expanding the spacetime metric around the Minkowski background as in Eq.~\eqref{lin-metric}, the quadratic gravitational action up to the order $\mathcal{O}(h^2)$ reads \cite{Biswas:2011ar}
\begin{equation}
\begin{array}{rl}
S=&\displaystyle \frac{1}{4}\int d^4x\left\lbrace \frac{1}{2}h_{\mu\nu}a(\Box)\Box h^{\mu\nu}-h_{\mu}^{\sigma}a(\Box)\partial_{\sigma}\partial_{\nu}h^{\mu\nu}\right.\\[4mm]
& \,\,\,\,\,\,\,\,\,\,\,\,\,\,\,\,\,\,\,\displaystyle +hc(\Box)\partial_{\mu}\partial_{\nu}h^{\mu\nu} -\frac{1}{2}hc(\Box)\Box h\\[4mm]
&\,\,\,\,\,\,\,\,\,\,\,\,\,\,\,\,\,\,\,\displaystyle\left.+\frac{1}{2}h^{\lambda\sigma}\frac{a(\Box)-c(\Box)}{\Box}\partial_{\lambda}\partial_{\sigma}\partial_{\mu}\partial_{\nu}h^{\mu\nu}\right\rbrace ,
\label{lin-quad-action}
\end{array}
\end{equation}
where $h\equiv\eta_{\mu\nu}h^{\mu\nu}$ and
\begin{equation}
\begin{array}{rl}
a(\Box)=& \displaystyle  1+\frac{1}{2}\mathcal{F}_2(\Box)\Box\,,\\[4mm]
c(\Box)=& \displaystyle 1-2\mathcal{F}_1(\Box)\Box-\frac{1}{2}\mathcal{F}_2(\Box)\Box\,.
\end{array}
\end{equation}
The related linearized field equations are represented by
\begin{equation}
\begin{array}{ll}
\displaystyle a(\Box)\left(\Box h_{\mu\nu}-\partial_{\sigma}\partial_{\nu}h_{\mu}^{\sigma}-\partial_{\sigma}\partial_{\mu}h_{\nu}^{\sigma}\right)+c(\Box)\left(\eta_{\mu\nu}\partial_{\rho}\partial_{\sigma}h^{\rho\sigma}\right.&\\[4mm]
\!\!\!\!\!\displaystyle +\left.\partial_{\mu}\partial_{\nu}h-\eta_{\mu\nu}\Box h\right)+ \frac{a(\Box)-c(\Box)}{\Box}\partial_{\mu}\partial_{\nu}\partial_{\rho}\partial_{\sigma}h^{\rho\sigma}=-2\kappa^2 T_{\mu\nu},&
\label{lin-field-eq}
\end{array}
\end{equation}
where 
\begin{equation}
T_{\mu\nu}=-\frac{2}{\sqrt{-g}}\frac{\delta S_m}{\delta g^{\mu\nu}}\,,
\end{equation}
is the stress-energy tensor generating the gravitational field, with $S_m$ being the matter action.

We are interested in finding the expression for the linearized metric generated by a static point-like source\footnote{Note that the linearized metric in Eq.~\eqref{isotr-metric} is expressed in isotropic coordinates, where $dr^2+r^2d\Omega^2=dx^2+dy^2+dz^2.$ }:
\begin{equation}
ds^2=-(1+2\Phi)dt^2+(1-2\Psi)(dr^2+r^2d\Omega^2),\label{isotr-metric}
\end{equation}
where $\Phi$ and $\Psi$ are the metric potentials generated by
\begin{equation}
T_{\mu\nu}=m\delta_{\mu}^0\delta_{\nu}^0\delta^{(3)}(\vec{r}).
\end{equation}
By using $\kappa h_{00}=-2\Phi$, $\kappa h_{ij}=-2\Psi\delta_{ij}$, $\kappa h=2(\Phi-3\Psi)$ and assuming the source to be static, that is $\Box\simeq \nabla^2,$ $T=\eta_{\rho\sigma}T^{\rho\sigma}\simeq-T_{00}=-\rho\,$, the field equations for the two metric potentials read\footnote{In order to obtain the differential equations in Eq.~\eqref{field-eq-pot}, we have considered and combined the trace and $(00)$-component of the field equations in Eq.~\eqref{lin-field-eq}.}
\begin{equation}
\begin{array}{rl}
\displaystyle \frac{a(a-3c)}{a-2c}\nabla^2\Phi(r)=&8\pi G\rho(r),\\[4mm]
\displaystyle \frac{a(a-3c)}{c}\nabla^2\Psi(r)=&- 8\pi G\rho(r),
\end{array}\label{field-eq-pot}
\end{equation}
where $a\equiv a(\nabla^2),$ $c\equiv c(\nabla^2)$ and $\rho(r)=m\delta^{(3)}(\vec{r})$. 

We can solve the two differential equations in Eq.~\eqref{field-eq-pot} by employing Fourier transform and then anti-transform to coordinate space. Thus, we obtain
\begin{equation}
\begin{array}{rl}
\Phi(r)=& \displaystyle -8\pi Gm\int \frac{d^3k}{(2\pi)^3}\frac{1}{k^2}\frac{a-2c}{a(a-3c)}e^{i\vec{k}\cdot \vec{r}}\\[4mm]
=& \displaystyle-\frac{4Gm}{\pi r}\int_0^{\infty}dk\frac{a-2c}{a(a-3c)}\frac{{\rm sin}(kr)}{k}\,,\\[4mm]
\Psi(r)=& \displaystyle 8\pi Gm\int \frac{d^3k}{(2\pi)^3}\frac{1}{k^2}\frac{c}{a(a-3c)}e^{i\vec{k}\cdot \vec{r}}\\[4mm]
=& \displaystyle\frac{4Gm}{\pi r}\int_0^{\infty}dk\frac{c}{a(a-3c)}\frac{{\rm sin}(kr)}{k}\,,
\end{array}\label{fourier-pot}
\end{equation}

where $a\equiv a(k^2)$ and $c\equiv c(k^2)$.

It is immediate to observe that, if $a=c$, the two metric potentials coincide, $\Phi=\Psi$. Therefore, as a special case we recover general relativity
\begin{equation}
a=c=1\,\,\Longrightarrow\,\,\Phi(r)=\Psi(r)=-\frac{Gm}{r}\,,
\end{equation}
as expected.

\section{Massless scalar field in a curved background and the Casimir effect}

\begin{figure}[t!]
\centering
\includegraphics[scale=.6]{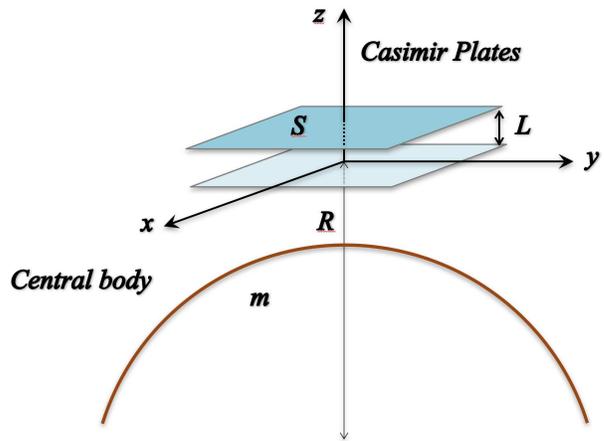}
\caption{In this picture, the configuration of the Casimir-like system in a gravitational field is represented.}
\label{fig_1}
\end{figure}

We now want to study the behavior of a massless scalar field $\psi(t,\,\vec{r})$ confined between two plates and embedded in a gravitational field (see Fig~\ref{fig_1}); to this aim, we basically follow the procedure presented in Ref.~\cite{Lambiase:2016bjy}. 

In the configuration of Fig.~\ref{fig_1}, $\psi(t,\,\vec{r})$ obeys the following field equation \cite{BirDav}:

\begin{equation}\label{KG_Massless_Eq}
\begin{array}{rl}
\displaystyle (\square +\xi \,{\cal R}) \psi(t,\,\vec{r})\,=& \displaystyle\,\frac{1}{\sqrt{-g}}\partial_{\alpha}\bigl[\sqrt{-g}\,g^{\alpha\beta}\,\partial_{\beta}\psi(t,\,\vec{r})\bigr]\\[4mm]
& \displaystyle+ \xi \,{\cal R}\,\psi(t,\,\vec{r})\,=\,0\,,
\end{array}
\end{equation}
where $\xi$ is the coupling parameter between geometry and matter\footnote{It is worth noting that higher-order couplings in the curvature between matter and geometry can in principle be inserted in the analysis, but since we are dealing with the linearized regime only the contribution $\mathcal{R}\,\psi$ survives in the field equations, which means that in the action we only consider the non-minimal coupling $\mathcal{R}\,\psi^2.$}. For simplicity, we consider the massless scalar field $\psi(t,\,\vec{r})$ confined between two parallel plates separated by a distance $L$ and with extension $S$, placed at a distance $R$ from the gravitational non-rotating source ($R\gg L, \sqrt{S}$).

We select a reference frame with the origin in the point-like source of gravity and the z-axis along the radial direction, perpendicular to the surface of the plates. We can then expand the metric tensor components and ${\cal R}(\vec{r})$ around the distance $R$ along the $z$ direction by using the gravitational potentials $\Phi(\vec{r})$ and $\Psi(\vec{r})$ given in Eq.~(\ref {fourier-pot})
\begin{eqnarray}\label{Expantion_Potentials}
\begin{array}{ll}
g_{00}(\vec{r})\simeq  -1-2\,\Phi_{0} -2\,\Phi_{1}\,z\,,
\\\\
g_{ij}(\vec{r})\simeq 1-2\,\Psi_{0} -2\,\Psi_{1}\,z\,,
\\\\
{\cal R}(\vec{r})\simeq {\cal R}_{0}+{\cal R}_{1}\,z\,,
\end{array}
\end{eqnarray}
where
\begin{eqnarray}\label{Expantion_Potentials2}
\nonumber
\Phi_{0}\,&=&\,\Phi(R), \,\,\,\,\,\,\, \Phi_{1}\,=\,\frac{d \Phi(r)}{dr}\Big|_{r=R}\,,\\
\label{Series}
\Psi_{0}\,&=&\,\Psi(R), \,\,\,\,\,\,\, \Psi_{1}\,=\,\frac{d \Psi(r)}{dr}\Big|_{r=R}\,,\\
\nonumber
{\cal R}_{0}\,&=&\,{\cal R}(R), \,\,\,\,\,\,\, {\cal R}_{1}\,=\,\frac{d {\cal R}(r)}{dr}\Big|_{r=R}\,,
\end{eqnarray}
and the variable $z$ is free to range in the interval $[0,L]$.
 
Adopting the metric Eq.~(\ref{Expantion_Potentials}), the field equation for the scalar field $\psi(t,\,\vec{r})$ of Eq.~(\ref{KG_Massless_Eq}) becomes
\begin{eqnarray}\nonumber
{\ddot \psi}(t,\,\vec{r})&-&\biggl[1+4\eta+ 4\gamma z \biggr] \nabla^2 \psi(t,\,\vec{r})\\\label{KG_Massless_Eq_Expantion}
&+&\xi\,\biggl[{\cal R}_{0}+{\cal R}_{1} z\biggl] \psi(t,\,\vec{r})=\,0\,,
\end{eqnarray}
where the dot indicates a derivative with respect to $t$, and
\begin{eqnarray}
\begin{array}{ll}
\eta\equiv\Phi_{0}+\Psi_{0}\,,
\\\\
\gamma\equiv\Phi_{1}+\Psi_{1}\,.
\end{array}
\end{eqnarray}
In order to solve Eq.~(\ref{KG_Massless_Eq_Expantion}) we perform the ansatz
\begin{eqnarray}\label{Ciccio}
\psi(t,\,\vec{r})\,=\,N\,\Upsilon(z)\,e^{i\,(\omega t-\vec{k}_\bot\cdot\vec{r}_\perp)}\,,
\end{eqnarray}
where $\vec{k}_\bot\equiv(k_x,k_y)$, $\vec{r}_\bot\equiv(x,y)$ and $N$ is a normalization constant. The field equation in Eq.~(\ref{KG_Massless_Eq_Expantion}) thus becomes
\begin{eqnarray}\label{KG_Massless_Eq_Expantion2}
(\partial^2_\chi+\chi)\,\Upsilon(\chi)\,=\,0\,,
\end{eqnarray}
where
\begin{eqnarray}\label{theta_zita}
\begin{array}{ll}
\chi\equiv\chi(z)\,=\,a-b\,z\equiv\,\alpha\,\beta^{-2/3}-\beta^{1/3}\,z\,,
\\\\
\alpha\,=\,\bigl( 1-4\,\eta\bigr)\omega^2-\vec{k}_\bot^2-\xi\,{\cal R}_0\,,
\\\\
\beta\,=\,4\,\omega^2\,\gamma+\xi\,{\cal R}_1\,.
\end{array}
\end{eqnarray}
The solution of Eq.~(\ref{KG_Massless_Eq_Expantion2}) can be expressed in terms of Bessel functions
\begin{equation}\label{Bessel_Equation}
\Upsilon(\chi)\,=\,\sqrt{\chi}\biggl[ C_1\,J_{1/3}\biggl( \frac{2}{3}\chi^{3/2}\biggr)+C_2\,J_{-1/3}\biggl( \frac{2}{3}\chi^{3/2}\biggr) \biggr]\,,
\end{equation}
where $C_1$ and $C_2$ are constants. We note that $\chi\gg 1$ since $\beta\ll\alpha$. For this reason, we can rewrite Eq.~(\ref{Bessel_Equation}) as~\cite{grad}
\begin{eqnarray}\label{Bessel_Equation_asymptotic}
\Upsilon(\chi)\,\simeq\,\sqrt{\frac{3}{\pi\,\sqrt{\chi}}}\sin\biggl[ \frac{2}{3}\chi^{3/2}+\tau \biggr]\,.
\end{eqnarray}
If we assume that the field $\psi$ satisfies Dirichlet boundary conditions on the plates, that is
\begin{eqnarray}\label{Dirichlet_boundary_conditions}
\psi(z=0)\,=\,\psi(z=L)\,=\,0 \,,
\end{eqnarray}
after some algebra, we find the relation
\begin{eqnarray}\label{Dirichlet_Relation}
\frac{2}{3}\biggl[ \chi^{3/2}\bigl( 0\bigr)-\chi^{3/2}\bigl( L\bigr)\biggr]\,=\,n\,\pi,
\end{eqnarray}
with $n$ being an integer number. 

From Eq.~(\ref{Dirichlet_Relation}), we compute the energy spectrum, which turns out to be
\begin{equation}\label{Energy_Spectrum}
\omega^2_n\,=\,\bigl(1+4\eta+2\gamma\,L\bigr)\biggl[\vec{k}_\bot^2+\biggl( \frac{n\pi}{L}\biggr)^2\biggr]+\xi\,\biggl[ {\cal R}_0+\frac{1}{2}{\cal R}_1\,L\biggr]\,.
\end{equation}
Finally, we derive the normalization constant
\begin{eqnarray}\label{N^2}
N_n^2\,=\,\frac{a\,b}{3\,S\,\omega_n\,n\bigl( 1-\Phi_0\bigl)}\,,
\end{eqnarray}
by means of the scalar product of QFT in curved spacetimes~\cite{BirDav}
\begin{equation}\label{scalar_product}
\begin{array}{rl}
\bigl(\psi_n,\psi_m\bigl)=&\displaystyle -i\int_V\sqrt{g_\Sigma}\,n^\mu\, dx\,dy\,dz\\[4mm]
& \displaystyle \times \biggl[\bigl(\partial_{\mu}\,\psi_n\bigr)\psi^*_m-\psi_n\bigl(\partial_{\mu}\,\psi_m\bigr)^*\biggr]\,.
\end{array}
\end{equation}
Note that $g_\Sigma$ is the induced metric on a spacelike Cauchy hypersurface $\Sigma$.

\subsection{Vacuum density energy}

In order to calculate the mean vacuum energy density ${\cal E}$ between the plates, we use the general relation \cite{BirDav}
\begin{equation}
 {\cal E}\,=\,\frac{1}{V_P}\sum_n\int d^2\,\vec{k}_\bot\int dx\,dy\,dz\,\sqrt{g_\Sigma} \,\bigl( g_{00}\bigr)^{-1}\,T_{00}\bigl( \psi_n,\psi_n^*\bigr)\,,
\end{equation}
where
\begin{equation}\label{Proper_Volume_Energy_Momentum_Tensor}
\begin{array}{rl}
V_P=& \displaystyle \int dx\,dy\,dz\,\sqrt{g_\Sigma}\,\simeq \, S\,L\biggl[1-3\Psi_0-\frac{3}{2}\Psi_{1} L\biggl]\,,\\[4mm]
\nonumber
T_{\mu\nu}=& \displaystyle \partial_{\mu}\,\psi\,\partial_{\nu}\,\psi-\frac{1}{2}g_{\mu\nu}g^{\alpha\beta}\partial_{\alpha}\,\psi\,\partial_{\beta}\,\psi\,,
\end{array}
\end{equation}
with $V_P$ being the proper volume and $T_{00}\bigl( \psi_n,\psi_n^*\bigr)$ a component of the energy-momentum tensor. 

Using the Schwinger proper-time representation and the $\zeta$-function regularization, we find the following expression for the mean vacuum energy density \cite{Lambiase:2016bjy}:
\begin{equation}\label{Energy_Casimir_2_Order}
\begin{array}{rl}
{\cal E}=& \displaystyle -\biggl[1+3\bigl(\Phi_0-\Psi_0\bigr) -\bigl( 2\Psi_{1}-\Phi_{1}\bigr) L_P\biggr]\frac{\pi^2}{1440\,L^4_P}\\[4mm]
&\displaystyle  +\frac{\xi\,{\cal R}_1}{192\,L_P}\,,
\end{array}
\end{equation}
where $L_P$ is the proper length of the cavity, defined as
\begin{equation}
L_P=\int dz\,\sqrt{g_{33}}\,\simeq\,L\left[1-\Psi_0-\frac{1}{2}\,\Psi_1\,L\right]\,.
\end{equation}

\subsection{Pressure}

The relevant physical observable of the Casimir effect is the attractive force between the plates, defined by ${\cal F}=-\frac{\partial E}{\partial L_P}$, where $E={\cal E}\,V_P$ is the Casimir vacuum energy. From Eqs.~\eqref{Expantion_Potentials}, one gets\footnote{The same computation was made in Ref.~\cite{Lambiase:2016bjy}, but the definition of proper area was not correctly taken into account. Here, we exhibit the correct expression for the force and the pressure. Hence, Eqs.~$(30)$, $(31)$ and $(32)$ in Ref.~\cite{Lambiase:2016bjy} need to be substituted by Eqs.~(\ref{Force_Casimir}), (\ref{Pressure_Casimir}) and (\ref{core}) of this paper, respectively.}

\begin{equation}\label{Force_Casimir}
{\cal F}\,=\,-\biggl[1+3\bigl(\Phi_0-\Psi_0\bigr) -\frac{2}{3}\left(2\Psi_{1}-\Phi_{1}\right) L_P\biggr]\frac{\pi^2\,S_P}{480\,L^4_P}\,,
\end{equation}
with $S_P$ being the proper area, defined as
\begin{equation}
S_P=\int dx\,dy\,\sqrt{g_{11}g_{22}}\,\simeq\,S\left[1-2\Psi_0-\,\Psi_1\,L\right]\,.
\end{equation}
By looking at Eq.~(\ref{Force_Casimir}), we can notice that the corrections related to the curvature do not contribute to $\mathcal{F}$. 

If now we introduce the Casimir pressure as ${\cal P}={\cal F}/S_p$, we finally obtain
\begin{equation}
\begin{array}{rl}
{\cal P}=& \displaystyle {\cal P}_0+{\cal P}_{G}\,,\\[2mm]
{\cal P}_0=&\displaystyle-\frac{\pi^2\,}{480\,L^4_P}\,,\\[2mm]
{\cal P}_{G}=&\displaystyle\biggl[3\bigl(\Phi_0-\Psi_0\bigr) -\frac{2}{3}\left(2\Psi_{1}-\Phi_{1}\right) L_P\biggr]{\cal P}_0\,,
\end{array}
\label{Pressure_Casimir}
\end{equation}
where ${\cal P}_0$ is the Casimir pressure in the flat case, while ${\cal P}_{G}$ is the correction induced by gravity. 

It must be emphasized that the quantity $\mathcal{P}_G$ is the sum of two distinct contributions, namely
\begin{equation}
\mathcal{P}_G=\mathcal{P}_{GR}+\mathcal{P}_Q\,,
\end{equation}
with $\mathcal{P}_{GR}$ being the contribution of Einstein's general relativity and $\mathcal{P}_Q$ the one arising from the quadratic part of the gravitational action in Eq. \eqref{quad-action}.

The relations in Eqs.~(\ref{Force_Casimir}) and (\ref{Pressure_Casimir}) give the corrections to the Casimir pressure up to the second order $\mathcal{O}(m/R^2)$. Differently from the case of GR (where contributions to the Casimir pressure appear only at the second order), for quadratic theories of gravity one obtains corrections already at $\mathcal{O}(m/R)$ provided that $\Phi_0\neq\Psi_0.$ This is an interesting outcome, since this contribution allows us to discriminate between GR and extended models of gravity. Furthermore, we also want to remark that the first-order correction increases the Casimir pressure, which is instead decreased by the second-order one.

\subsection{Experimental constraints}

The final step of our analysis consists in constraining quadratic theories of gravity by means of the present experimental sensitivity. 

Since we can impose the bound $|{\cal P}_{G}|\lesssim \delta {\cal P}$, where $\delta {\cal P}$ is the experimental error (as firstly discussed in Ref.~\cite{Lambiase:2016bjy}), we then obtain
\begin{eqnarray}\label{core}
\biggl|3\bigl(\Phi_0-\Psi_0\bigr) -\frac{2}{3}\left(2\Psi_{1}-\Phi_{1}\right) L_P\biggr|\lesssim \frac{\delta {\cal P}}{{\cal P}_0}\,,
\end{eqnarray}
that gives the bound for the free parameters in the context of quadratic theories of gravity. We want to stress once again that, for theories whose potentials satisfy the property $\Phi_0\neq\Psi_0,$ it is possible to immediately deduce a constraint on the extra parameters, since Einstein's general relativity gives no contribution at the first order in the corrections.

\section{Application to several quadratic theories of gravity}

Let us now apply the above formalism of the Casimir effect to several quadratic theories of gravity. We consider both local and non-local models and compute the corresponding metric potentials and the linearized Ricci scalar. Their knowledge is indispensable to study the dynamics of a scalar field between two plates in a slightly curved background, and thus to constrain the parameters of new physics appearing in such extended theories. The Ricci scalar does not appear in the formula for the pressure (see Eq.~\eqref{Force_Casimir}), but we compute it for the sake of completeness.

Note that the second term in the l.h.s. of Eq.~\eqref{core} is of order $\mathcal{O}\left(m/R^2\right)$ for every analyzed theory, whereas the first one goes like $\mathcal{O}\left(m/R\right)$. Since we assume $R$ to be large, when $\Phi\neq \Psi,$ we can safely neglect the second-order contributions\footnote{These factors are also multiplied by $L_P$, which makes our assumption even stronger, given that the proper length of the cavity is small.} that stem from $\Psi_1$ and $\Phi_1.$ 

We also want to stress that, in the case of Einstein's GR, $\Phi=\Psi=-Gm/r,$ and thus we recover the results for the Casimir energy and the Casimir pressure obtained in Ref.~\cite{Sorge}.

\subsection{$f(\mathcal{R})$-gravity}

We first address the easiest extension of the Einstein-Hilbert action by including a Ricci squared contribution with a constant form-factor
\begin{equation}
\mathcal{F}_1=\alpha\,,\,\,\,\mathcal{F}_2=0\,\,\Longrightarrow\,\,a=1\,,\,\,\,c=1-2\,\alpha\,\Box\,.
\end{equation}
This choice belongs to the class of $f(\mathcal{R})$ theories, where the Lagrangian is truncated up to the order $\mathcal{O}(\mathcal{R}^2)$
\begin{equation}
f(\mathcal{R})\simeq \mathcal{R}+\alpha \mathcal{R}^2,
\end{equation}
and where the cosmological constant is set to zero.

For the above selection of the form-factors, the two metric potentials in Eq.~\eqref{fourier-pot} become
\begin{equation}
\begin{array}{rl}
\Phi(r)=&\displaystyle -\frac{Gm}{r}\left(1+\frac{1}{3}e^{-m_0r}\right),\\[4mm]
\Psi(r)=&\displaystyle -\frac{Gm}{r}\left(1-\frac{1}{3}e^{-m_0r}\right),
\end{array}
\end{equation}
where $m_0=1/\sqrt{3\,\alpha}$ is the mass of the spin-$0$ massive degree of freedom coming from the Ricci scalar squared contribution.

The linearized Ricci scalar is given by
\begin{equation}
\mathcal{R}=\frac{2\,G\,m}{r}\,e^{-m_0r}\,m_0^2\,.\label{ricci-f(R)}
\end{equation}
Plugging the above expressions in Eq.~\eqref{core}, we have
\begin{equation}\label{pfr}
e^{-m_0R}\lesssim\frac{\delta \mathcal{P}}{\mathcal{P}_0}\frac{R}{2Gm}\,.
\end{equation}

\subsection{Stelle's fourth order gravity}

Let us now consider Stelle's fourth order gravity~\cite{-K.-S.}, which is achieved with the following form-factors:
\begin{equation}
\mathcal{F}_1=\alpha\,,\,\mathcal{F}_2=\beta\,\Longrightarrow\,a=1+\frac{1}{2}\beta\,\Box,\,\,c=1-2\alpha\,\Box-\frac{1}{2}\beta\,\Box\,.
\end{equation}
Differently from the previous case, the Ricci tensor squared contribution in the action is clearly recognizable through a constant, non-vanishing form-factor. It is possible to check that the gravitational action related to this model turns out to be renormalizable~\cite{-K.-S.}. 

For the above choice of the form-factors, the two metric potentials in Eq.~\eqref{fourier-pot} now read
\begin{equation}
\begin{array}{rl}
\Phi(r)=&\displaystyle -\frac{Gm}{r}\left(1+\frac{1}{3}e^{-m_0r}-\frac{4}{3}e^{-m_2r}\right),\\[4mm]
\Psi(r)=&\displaystyle -\frac{Gm}{r}\left(1-\frac{1}{3}e^{-m_0r}-\frac{2}{3}e^{-m_2r}\right),\label{stelle-pot}
\end{array}
\end{equation}
where $m_0=2/\sqrt{12\,\alpha+\beta}$ and $m_2=\sqrt{2/(-\beta)}$ correspond to the masses of the spin-$0$ and of the spin-$2$ massive mode, respectively. 

In order to avoid tachyonic solutions, we need to require $\beta<0$. In addition to that, the spin-$2$ mode is a ghost-like degree of freedom. Such an outcome is not surprising, since it is known that, for any local higher derivative theory of gravity, ghost-like degrees of freedom always appear~\cite{Asorey:1996hz}.

The linearized Ricci scalar generated by a point-like source and derived from the metric with potentials given by Eq.~\eqref{stelle-pot} is the same as the one of Eq.~\eqref{ricci-f(R)}
\begin{equation}
\mathcal{R}=\frac{2\,G\,m}{r}\,e^{-m_0r}\,m_0^2\,.\label{ricci-stelle}
\end{equation}
It is opportune to explain why the linearized Ricci scalar for Stelle's theory has the same form of the linearized Ricci scalar in Eq.~\eqref{ricci-f(R)}. Given the following two metric potentials:
\begin{equation}
\begin{array}{rl}
\Phi(r)=&\displaystyle -\frac{Gm}{r}\left(1+a\,e^{-m_0r}+b\,e^{-m_2r}\right),\\[4mm]
\Psi(r)=&\displaystyle -\frac{Gm}{r}\left(1+c\,e^{-m_0r}+d\,e^{-m_2r}\right),
\end{array}
\end{equation}
the related linearized Ricci scalar is
\begin{equation}
\mathcal{R}=\frac{2Gm}{r}\left((a-2c)\,e^{-m_0r}\,m_0^2+(b-2d)\,e^{-m_2r}\,m_2^2\right).
\end{equation}
We can now understand that, in the case of Stelle's theory, the linearized Ricci scalar is given by Eq.~\eqref{ricci-stelle}, because $b-2\,d=-4/3-2\,(-2/3)=0$. Thus, only the term depending on $m_0$ survives.
We also want to stress that the curvature invariant in Stelle's theory is still singular at $r=0.$

The correction to the Casimir pressure $\mathcal{P}$ given by this model is 
\begin{equation}\label{stellec} 
\left|e^{-m_0R}-e^{-m_2R}\right|\lesssim\frac{\delta \mathcal{P}}{\mathcal{P}_0}\frac{R}{2Gm}\,.
\end{equation}
As expected, for $m_2\rightarrow \infty\,\,(\beta\rightarrow 0)$ we immediately obtain the same expression of Eq.~(\ref{pfr}).

\subsection{Sixth order gravity}
Here, we deal with a sixth order gravity model, which is an example of super-renormalizable theory~\cite{Asorey:1996hz,Giacchini:2018gxp,Accioly:2016qeb}
\begin{equation}
\begin{array}{ll}
\mathcal{F}_1=\alpha\,\Box\,,\,\,\mathcal{F}_2=\beta\,\Box&\\ 
\displaystyle \,\,\,\,\,\,\,\,\,\,\,\,\,\,\,\,\,\,\,\,\,\Longrightarrow\,a=1+\frac{1}{2}\beta\,\Box^2,\,\,\,c=1-2\alpha\Box^2-\frac{1}{2}\beta \Box^2\,.&
\end{array}
\end{equation}
The form-factors of the two metric potentials in Eq.~\eqref{fourier-pot} assume the following expression:
\begin{equation}
\!\!\!\!\begin{array}{rl}
\Phi(r)\!=&\!\!\!\displaystyle -\frac{Gm}{r}\!\left(1+\frac{1}{3}e^{-m_0r}\,{\rm cos}(m_0r)-\frac{4}{3}e^{-m_2r}\,{\rm cos}(m_2r)\right)\!,\\[4mm]
\Psi(r)\!=&\!\!\!\displaystyle -\frac{Gm}{r}\!\left(1-\frac{1}{3}e^{-m_0r}\,{\rm cos}(m_0r)-\frac{2}{3}e^{-m_2r}\,{\rm cos}(m_2r)\right)\!,\label{sixth-pot}
\end{array}
\end{equation}
where the masses of the spin-$0$ and spin-$2$ degrees of freedom are now given by $m_0=2^{-1/2}(-3\,\alpha-\beta)^{-1/4}$ and $m_2=(2\,\beta)^{-1/4}$, respectively.

This time, tachyonic solutions are avoided for $-3\,\alpha-\,\beta>0$, which can be satisfied by the requirement $\alpha<0$ and $-3\,\alpha>\,\beta$, with $\beta>0$. The current higher derivative theory of gravity has no real ghost-modes around the Minkowski background but a pair of complex conjugate poles with equal real and imaginary parts~\cite{Accioly:2016qeb}, and corresponds to the so called Lee-Wick gravity~\cite{Modesto:2015ozb}. It is worthwhile noting that in this higher derivative theory the unitarity condition is not violated, indeed the optical theorem still holds~\cite{Anselmi,Anselmi:2017yux,Anselmi:2017lia}.

The linearized Ricci scalar for this model turns out to be
\begin{equation}
\mathcal{R}=\frac{2\,G\,m}{r}\,{\rm sin}(m_0r)\,e^{-m_0r}\,m_0^2\,.
\end{equation}
We remark that, in the linearized Ricci scalar, there is no $m_2$ dependence for the same reason explained before. However, the crucial difference here is that the curvature is regular at $r=0$.

In this case, the bound on the pressure in Eq.~\eqref{core} becomes 
\begin{equation}\label{psog}
\left|e^{-m_0R}\mathrm{cos}\left(m_0R\right)-e^{-m_2R}\mathrm{cos}\left(m_2R\right)\right|\lesssim\frac{\delta \mathcal{P}}{\mathcal{P}_0}\frac{R}{2Gm}\,.
\end{equation}
This expression resembles the one of Eq.~\eqref{stellec}, apart from oscillatory functions of the products $m_0R$ and $m_2R$. Furthermore, as for the case of Stelle's theory, in Eq.~\eqref{psog} the limit $m_2\rightarrow \infty$ smoothly approaches the outcome of Eq.~(\ref{pfr}), with the only difference that now the periodic function $\mathrm{cos}(m_0R)$ is in principle not negligible.

\subsection{Ghost-free infinite derivative gravity}

We now wish to consider an example of ghost-free non-local theory of gravity~\cite{Biswas:2005qr,Modesto:2011kw,Biswas:2011ar,Biswas:2013cha,Biswas:2016etb,Edholm:2016hbt,Buoninfante,Koshelev:2017bxd,Buoninfante:2018xiw,Buoninfante:2018rlq,Buoninfante:2018stt,Buoninfante:2018mre,Buoninfante:2018xif,Mazumdar:2018xjz}. For the sake of clarity, we decide to adopt the simplest ghost-free choice for the non-local form-factors~\cite{Biswas:2011ar}
\begin{equation}
\mathcal{F}_1=-\frac{1}{2}\mathcal{F}_2=\,\frac{1-e^{-\Box/M_s^2}}{2\,\Box}\,\,\Longrightarrow\,\,a=c=e^{-\Box/M_s^2}, \label{ghost-free-choice}
\end{equation}
where $M_s$ is the scale at which the non-locality of the gravitational interaction should become manifest. Note that, for the special ghost-free choice in Eq.~\eqref{ghost-free-choice}, {\it no} extra degrees of freedom other than the massless transverse spin-$2$ graviton propagate around the Minkowski background. 

Since we have chosen $a=c$, the metric potentials of Eq.~\eqref{fourier-pot} coincide
\begin{equation}
\Phi(r)=\Psi(r)=-\frac{Gm}{r}\,{\rm Erf}\left(\frac{M_s\,r}{2}\right)\label{ghost-free-pot}\,,
\end{equation}
The corresponding linearized Ricci scalar is given by
\begin{equation}
\mathcal{R}=\frac{G\,m\,M_s^3\,e^{-\frac{M_s^2r^2}{4}}}{\sqrt{\pi}},
\end{equation}
which shows a Gaussian behavior induced by the presence of non-local gravitational interaction, whose role is to smear out the point-like source at the origin.

Since the two metric potentials coincide, we get no contribution at the first order, but we can constrain such a theory starting from the second-order correction. If this is the case, then the equation for the constraint reads
\begin{equation}\label{pidg}
\left|\frac{e^{-\frac{M_s^2R^2}{4}}M_s}{\sqrt{\pi}}-\frac{1}{R}\mathrm{Erf}\left(\frac{M_sR}{2}\right)\right|\lesssim\frac{\delta \mathcal{P}}{L_P\mathcal{P}_0}\frac{3R}{2\,Gm}\,.
\end{equation}
It is worth emphasizing that it would be very difficult to constrain this kind of non-local theory with Solar system's experiments, since the non-locality of the gravitational interaction described by the metric with the potential in Eq.~(\ref{ghost-free-pot}) is expected to become relevant only at scales comparable to the Schwarzschild radius \cite{Koshelev:2017bxd,Buoninfante:2018xiw,Buoninfante:2018rlq}.

\subsection{Non-local gravity with non-analytic form-factors }

For the last case, we consider two models of non-local infrared extension of Einstein's general relativity, where form-factors are non-analytic functions of $\Box$. These theories are inspired by quantum corrections to the effective action of quantum gravity~\cite{Bravinsky,Deser:2007jk,Deser:2013uya,Belgacem:2017cqo,Tan:2018bfp,Conroy:2014eja,Woodard:2018gfj}.
\begin{enumerate}
	\item The first model is described by the following choice of the form-factors:
	\begin{equation}
	\mathcal{F}_1=\frac{\alpha}{\Box}\,,\,\,\,\,\mathcal{F}_2=0\,\,\,\Longrightarrow\,\,a=1\,,\,\,\,\,c=1-2\alpha\,.\label{non-local-choice1}
	\end{equation}
	The two metric potentials are infrared modifications of the Newtonian one
	\begin{equation}
	\begin{array}{rl}
	\Phi(r)=&\displaystyle -\frac{Gm}{r}\left(\frac{4\alpha-1}{3\alpha-1}\right)\,,\\[4mm]
	\Psi(r)=&\displaystyle -\frac{Gm}{r}\left(\frac{2\alpha-1}{3\alpha-1}\right)\,.
	\end{array}
	\end{equation}
    Such a spacetime metric has a vanishing linearized Ricci scalar, 
    \begin{equation}
    \mathcal{R}=0\,,
    \end{equation}
	but it is not Ricci-flat, since some components of $\mathcal{R}_{\mu\nu}$ are non-vanishing.
	
	The expression for the Casimir pressure correction given by such a quadratic theory of gravity leads to the bound
	\begin{equation}\label{pb.1}
	\left|\frac{\alpha}{3\alpha-1}\right|\lesssim\frac{\delta \mathcal{P}}{\mathcal{P}_0}\frac{R}{6Gm}\,.
	\end{equation}
	In the last Equation, if we treat $\alpha$ as a small parameter, we obtain the constraint
	\begin{equation}\label{pb1}
	\left|\alpha\right|\lesssim\frac{\delta \mathcal{P}}{\mathcal{P}_0}\frac{R}{6Gm}\,.
	\end{equation}
	Hence, differently from the other results, by performing such an expansion it is possible to have a direct access to the free parameter of the theoretical model, thus immediately obtaining a constraint on it.

	\item The non-local form factors for the second model are 
	\begin{equation}
	\mathcal{F}_1=\frac{\beta}{\Box^2}\,,\,\,\,\,\mathcal{F}_2=0\,\,\,\Longrightarrow\,\,a=1\,,\,\,\,\,c=1-\frac{2\beta}{\Box}\,.\label{non-local-choice2}
	\end{equation}
	In this framework, the infrared modification is not a constant, but the metric potentials show a Yukawa-like behavior
	\begin{equation}
	\begin{array}{rl}
	\Phi(r)=&\displaystyle -\frac{Gm}{r}\frac{4}{3}\left(1-\frac{1}{4}e^{-\sqrt{3\beta}r}\right),\\[4mm]
	\Psi(r)=&\displaystyle -\frac{Gm}{r}\frac{2}{3}\left(1+\frac{1}{2}e^{-\sqrt{3\beta}r}\right).
	\end{array}
	\end{equation}
	The corresponding linearized Ricci scalar is non-vanishing and negative
	\begin{equation}
	\mathcal{R}=-\frac{6\,G\,m}{r}\,e^{-\sqrt{3\beta}r}\,\beta\,.
	\end{equation}
	For this second example of non-local gravity with non-analytic form-factors, we obtain the bound
	\begin{equation}\label{pb2}
	\left|1-e^{-\sqrt{3\beta}R}\right|\lesssim\frac{\delta \mathcal{P}}{\mathcal{P}_0}\frac{R}{2Gm}\,.
	\end{equation}

\end{enumerate}

\section{Discussions and Conclusions}\label{conclus}

In this paper, we have discussed the Casimir effect in the context of quadratic theories of gravity. In particular, we have focused our attention on the corrections to the Casimir pressure which come from the gravitational sector that makes these models different from GR. We have also described the possibility to infer several constraints on the free parameters of the above theories in a completely transparent way, by relating them to the current experimental errors. In all the analyzed contexts, we have worked in the weak field regime, in which computations related to the metric potentials and to the Casimir effect are easy to carry out. This has proved to be convenient in order to immediately separate the Casimir pressure in all of its relevant contributions, as seen in Eq.~\eqref{Pressure_Casimir}.

A feature which is omnipresent in the outcomes of Eqs.~\eqref{pfr}, \eqref{psog}, \eqref{pidg}, \eqref{pb1} and \eqref{pb2} is that they all depend on the quantity $({\delta\mathcal{P}}/{\mathcal{P}_0})(R/Gm)$. If we consider the case of the Earth, and thus $R\equiv R_{\oplus}$, $m\equiv M_{\oplus}$, we get
\begin{equation}
\frac{\delta\mathcal{P}}{\mathcal{P}_0}\frac{R_{\oplus}}{GM_{\oplus}}\simeq 10^6\,,
\end{equation}
if we assume the value $\delta\mathcal{P}/\mathcal{P}_0\simeq 10^{-3}$, stemming from the current experimental precision~\cite{Moste}.

In order to enhance the attained bounds, there is the necessity either to implement the experiment of the Casimir effect near another astrophysical object (which is not the Earth) or to significantly improve the sensitivity of the experimental instruments. For what concerns the former, however, there is also the problem to adapt the solution of the Klein-Gordon equation to the nature of the spacetime that is under consideration. For instance, in the case of blackholes the ratio $R/Gm$ is substantially lowered, but our formalism based on a linear regime is no longer valid.

As a final observation, we want to focus the attention one more time on the fact that the first-order corrections to the pressure $\mathcal{P}$ in Eq.~\eqref{Pressure_Casimir} cannot be attributable to GR, since in Einstein's theory $\Phi_0=\Psi_0\,$. Hence, any gravitational contribution to the Casimir pressure arising at this order is intimately connected to an extended theory of gravity for which $a\neq c$ (see Eq. \eqref{fourier-pot}). 
From an experimental point of view, such a consideration would also imply a more stringent bound on the free parameters of these models. In this perspective, future experiments of the Casimir effect in curved background might represent a powerful tool to test and constrain extended theories of gravity.

\acknowledgements  The authors are thankful to Anupam Mazumdar. LB would like to thank Breno Giacchini and Shubham Maheshwari for discussions.


\begin{thebibliography}{99}
	
	\bibitem{-C.-M.}C. M. Will, 
	Living Rev. Rel. \textbf{17}, 4 (2014).
	
	\bibitem{-K.-S.}K. S. Stelle, 
	Phys. Rev. D \textbf{16}, 953 (1977); 
	K. S. Stelle, 
	Gen. Rel. Grav. \textbf{9}, 353--371 (1978).
	
	\bibitem{Anselmi} 
	D.~Anselmi, 
	JHEP {\bf 1706}, 086 (2017);
	
	\bibitem{Anselmi:2017ygm} 
	D.~Anselmi, 
	JHEP {\bf 1802}, 141 (2018); 
	D.~Anselmi and M.~Piva, 
	JHEP {\bf 1805}, 027 (2018); 
	D.~Anselmi and M.~Piva, 
	JHEP {\bf 1811}, 021 (2018); 
	D.~Anselmi, 
	arXiv:1809.05037 [hep-th].
	
	\bibitem{starobinski} A. A. Starobinski, 
	Phys. Lett. B \textbf{91}, 99--102 (1980); 
	A. A. Starobinski, 
	Proceedings of the Second Seminar
	``Quantum Theory of Gravity'', Moscow, 13-15 October 1981, INR Press, Moscow, 58-72, (1982); 
	A. A. Starobinsky, 
	Sov. Astron. Lett. \textbf{9}, 302 (1983).
	
	
	
	\bibitem{capoz1} S. Capozziello, G. Lambiase, M. Sakellariadou and An. Stabile, 
	Phys. Rev. D \textbf{91}, 044012 (2015).
	
	\bibitem{lamb1}G. Lambiase, M. Sakellariadou, A. Stabile and An. Stabile, 
	JCAP \textbf{1507}, 003 (2015).
	
	\bibitem{lamb2}G. Lambiase, M. Sakellariadou and A. Stabile, 
	JCAP \textbf{1312}, 020 (2013).
	
	\bibitem{radicella}N. Radicella, G. Lambiase, L. Parisi and G. Vilasi, 
	JCAP \textbf{1412}, 014 (2014).
	
	\bibitem{capoz3} S. Capozziello and G. Lambiase, 
	Int. J. Mod. Phys. D \textbf{12}, 843--852 (2003).
	
	\bibitem{calchi}S. Calchi Novati, S. Capozziello and G. Lambiase, 
	Grav. Cosmol. \textbf{6}, 173--180 (2000).
	
	\bibitem{capoz4}S. Capozziello, G. Lambiase and H. J. Schmidt, 
	Annalen Phys. \textbf{9}, 39--48 (2000).
	
	\bibitem{capoz5}S. Capozziello, G. Lambiase, G. Papini and G. Scarpetta, 
	Phys. Lett. A \textbf{254}, 11-17 (1999).
	
	
	\bibitem{Asorey:1996hz} 
	M.~Asorey, J.~L.~Lopez and I.~L.~Shapiro,
	Int.\ J.\ Mod.\ Phys.\ A {\bf 12}, 5711 (1997).
	
	\bibitem{stabile_stabile_cap}
	A. Stabile, An. Stabile and S. Capozziello, 
	Phys. \ Rev. \ D , {\bf 88}, 124011(9) (2013)
	
	\bibitem{stabstab} A. Stabile, An. Stabile,
	Phys. \ Rev. \ D , {\bf 85}, 044014 (2012)
	
	
	\bibitem{LamMohSta} G. Lambiase, S. Mohanty and An. Stabile,
	Eur. \ Phys.\ J.\ C  \textbf{78}, 350 (2018)
	
	\bibitem{Tomboulis:1997gg} 
	E.~T.~Tomboulis,
	hep-th/9702146.
	
	\bibitem{Biswas:2005qr} 
	T.~Biswas, A.~Mazumdar and W.~Siegel,
	JCAP {\bf 0603}, 009 (2006),
	
	\bibitem{Modesto:2011kw} 
	L.~Modesto,
	Phys.\ Rev.\ D {\bf 86}, 044005 (2012).
	
	\bibitem{Biswas:2011ar} 
	T.~Biswas, E.~Gerwick, T.~Koivisto and A.~Mazumdar,
	Phys.\ Rev.\ Lett.\  {\bf 108}, 031101 (2012).
	
	\bibitem{Biswas:2013cha} 
	T.~Biswas, A.~Conroy, A.~S.~Koshelev and A.~Mazumdar,
	Class.\ Quant.\ Grav.\  {\bf 31}, 015022 (2014);
	Erratum: Class.\ Quant.\ Grav.\  {\bf 31}, 159501 (2014).
	
	\bibitem{Biswas:2016etb} 
	T.~Biswas, A.~S.~Koshelev and A.~Mazumdar,
	Fundam.\ Theor.\ Phys.\  {\bf 183}, 97 (2016).
	T.~Biswas, A.~S.~Koshelev and A.~Mazumdar,
	Phys.\ Rev.\ D {\bf 95}, 043533 (2017).
	
	\bibitem{Koshelev:2017bxd} 
	A.~S.~Koshelev and A.~Mazumdar,
	Phys.\ Rev.\ D {\bf 96}, no. 8, 084069 (2017),
	
	\bibitem{casimir}
	H. Casimir, 
	Proc. K. Ned. Akad. Wet. \textbf{51} 793 (1948); 
	H. Casimir and D. Polder, 
	Phys. Rev. \textbf{73}, 360 (1948).
	
	\bibitem{flat}
	K. A. Milton, \emph{The Casimir effect: Physical Manifestations of Zero-Point Energy}, River edge: World Scientific, 2001; 
	V.V. Nesterenko, G. Lambiase and G. Scarpetta, 
	Riv. Nuovo Cim. \textbf{27}, N6, 1-74 (2004); 
	V.V. Nesterenko, G. Lambiase and G. Scarpetta, 
	Annals Phys. \textbf{298}, 403 (2002); 
	V.V. Nesterenko, G. Lambiase and G. Scarpetta, 
	Int. J. Mod. Phys. A \textbf{17}, 790 (2002); 
	V.V. Nesterenko, G. Lambiase and G. Scarpetta, 
	Phys. Rev. D \textbf{64}, 025013 (2001); 
	V.V. Nesterenko, G. Lambiase and G. Scarpetta, 
	J. Math. Phys. \textbf{42}, 1974 (2001); 
	G. Lambiase, G. Scarpetta and V.V. Nesterenko, 
	Mod. Phys. Lett. A \textbf{16}, 1983 (2001); 
	G. Lambiase, V.V. Nesterenko and M. Bordag, 
	J. Math. Phys. \textbf{40}, 6254 (1999); 
	M. Bordag, U. Mohideen and V. M. Mostepanenko, 
	Phys. Rep. \textbf{353}, 1 (2001);  
	C. Genet, A. Lambrecht and S. Reynaud, 
	\emph{On the Nature of Dark Energy}, 18th IAP Coll. on the Nature of Dark Energy: Observations and Theoretical Results in the Accelerating Universe, Paris, France, 1-5 July 2002, ed P. Brax, J. Martin, J. P. Uzan (Fronter Group), 121-30; 
	G. Bressi, G. Carugno, R. Onofrio and G. Ruoso, 
	Phys. Rev. Lett. \textbf{88}, 041804 (2002).
	
	\bibitem{Sorge}F. Sorge, 
	Class. Quantum Grav. \textbf{22}, 5109 (2005).
	
	\bibitem{curved}
	V.~G.~Bagrov, I.~L.~Buchbinder and S.~D.~Odintsov, 
	Phys.\ Lett.\ B {\bf 184}, 202 (1987) [Yad.\ Fiz.\  {\bf 45}, 1192 (1987)];
	I.~L.~Buchbinder, P.~M.~Lavrov and S.~D.~Odintsov, 
	Nucl.\ Phys.\ B {\bf 308}, 191 (1988);  
	S.~D.~Odintsov,
	Sov.\ Phys.\ J.\  {\bf 32}, 458 (1989); 
	M. R. Setare, 
	Class. Quantum Grav. \textbf{18}, 2097 (2001); 
	E. Calloni, L. di Fiore, G. Esposito, L. Milano and L. Rosa, 
	Int. J. Mod. Phys. A \textbf{17}, 804 (2002); 
	G. Esposito, G. M. Napolitano and L. Rosa, 
	Phys. Rev. D \textbf{77}, 105011 (2008); 
	G. Bimonte, G. Esposito and L. Rosa, 
	Phys. Rev. D \textbf{78}, 024010 (2008); 
	E. Calloni, M. De Laurentis, R. De Rosa, F. Garufi, L. Rosa, L. Di Fiore, G. Esposito, C. Rovelli, P. Ruggi, and F. Tafuri, 
	Phys. Rev. D \textbf{90}, 022002 (2014); B. Nazari, 
	Eur. Phys. J. C \textbf{75}, 501 (2015); 
	P. Bueno, P. A. Cano, V. S. Min and M. R. Visser, 
	Phys. Rev. D \textbf{95}, 044010 (2017); 
	M. R. Tanhayi and R. Pirmoradian, 
	R. Int. J. Theor. Phys. \textbf{55}, 766 (2016);  
	S. A. Fulling, K. A. Milton, P. Parashar, A. Romeo, K. V. Shajesh and J. Wagner, 
	Phys. Rev. D \textbf{76}, 025004 (2007); 
	K. A. Milton, P. Parashar, K. V. Shajesh and J. Wagner, 
	J. Phys. A \textbf{40}, 10935 (2007); 
	K. V. Shajesh, K. A. Milton, P. Parashar and J. A. Wagner, 
	J. Phys. A \textbf{41}, 164058 (2008); 
	K. A. Milton, K. V. Shajesh, S. A. Fulling and P. Parashar, 
	Phys. Rev. D \textbf{89}, 064027 (2014); 
	K. A. Milton, S. A. Fulling, P. Parashar, A. Romeo, K. V. Shajesh and J. A. Wagner, 
	J. Phys. A \textbf{41}, 164052 (2008).
	
	\bibitem{lv}
	M.~Blasone, G.~Lambiase, L.~Petruzziello and A.~Stabile,
	Eur.\ Phys.\ J.\ C {\bf 78}, no. 11, 976 (2018).
	
	\bibitem{lv1}
	M.~B.~Cruz, E.~R.~Bezerra de Mello and A.~Y.~Petrov,
	Phys.\ Rev.\ D {\bf 96}, no. 4, 045019 (2017);
	M.~B.~Cruz, E.~R.~Bezerra De Mello and A.~Y.~Petrov,
	Mod.\ Phys.\ Lett.\ A {\bf 33}, 1850115 (2018);
	M. Frank and I. Turan, 
	Phys. Rev. D \textbf{74}, 033016 (2006); 
	A. Martin-Ruiz and C. A. Escobar, 
	Phys. Rev. D \textbf{94}, 076010 (2016).
	
	\bibitem{mixed-fields}
	M. Blasone, G. G. Luciano, L. Petruzziello and L. Smaldone, 
	Phys. Lett. B {\bf 786}, 278 (2018).
	
	\bibitem{Lambiase:2016bjy} 
	G.~Lambiase, A.~Stabile and An.~Stabile,
	Phys.\ Rev.\ D {\bf 95}, 084019 (2017).
	
	\bibitem{BirDav} N. ~D. ~Birrell and P. ~C. ~W. ~Davies, \emph{Quantum Fields in Curved Space}, Cambridge: Cambridge University Press, 1982.
	
	\bibitem{grad}
	I.~S.~Gradshteyn and I.~M.~Ryzhik, \emph{Table of Integrals, Series
		and Products}, New York, Academic, 1980.
	
	\bibitem{Giacchini:2018gxp} 
	B.~L.~Giacchini and T.~de Paula Netto,
	arXiv:1806.05664 [gr-qc].
	
	\bibitem{Accioly:2016qeb} 
	A.~Accioly, B.~L.~Giacchini and I.~L.~Shapiro,
	Phys.\ Rev.\ D {\bf 96}, no. 10, 104004 (2017). B.~L.~Giacchini,
	Phys.\ Lett.\ B {\bf 766}, 306 (2017).
	
	\bibitem{Modesto:2015ozb} 
	L.~Modesto and I.~L.~Shapiro,
	Phys.\ Lett.\ B {\bf 755}, 279 (2016); L.~Modesto,
	Nucl.\ Phys.\ B {\bf 909}, 584 (2016).
	
	\bibitem{Anselmi:2017yux} 
	D.~Anselmi and M.~Piva,
	JHEP {\bf 1706}, 066 (2017).
	
	\bibitem{Anselmi:2017lia} 
	D.~Anselmi and M.~Piva, 
	Phys.\ Rev.\ D {\bf 96}, no. 4, 045009 (2017).
	
	
	\bibitem{Edholm:2016hbt} 
	J.~Edholm, A.~S.~Koshelev and A.~Mazumdar,
	Phys.\ Rev.\ D {\bf 94}, no. 10, 104033 (2016).
	
	
	\bibitem{Buoninfante} L. Buoninfante, Master's Thesis (2016),
	{[}arXiv:1610.08744v4 {[}gr-qc{]}{]}.
	
	\bibitem{Buoninfante:2018xiw} 
	L.~Buoninfante, A.~S.~Koshelev, G.~Lambiase and A.~Mazumdar,
	JCAP {\bf 1809} no.09, 034 (2018).
	
	\bibitem{Buoninfante:2018rlq} 
	L.~Buoninfante, A.~S.~Koshelev, G.~Lambiase, J.~Marto and A.~Mazumdar,
	JCAP {\bf 1806}, no. 06, 014 (2018).
	
	\bibitem{Buoninfante:2018stt}
	L.~Buoninfante, G.~Harmsen, S.~Maheshwari and A.~Mazumdar,
	Phys.\ Rev.\ D {\bf 98} (2018) no.8,  084009.
	
	\bibitem{Buoninfante:2018mre}
	L.~Buoninfante, G.~Lambiase and A.~Mazumdar,
	arXiv:1805.03559 [hep-th].
	
	\bibitem{Buoninfante:2018xif}
	L.~Buoninfante, A.~S.~Cornell, G.~Harmsen, A.~S.~Koshelev, G.~Lambiase, J.~Marto and A.~Mazumdar,
	Phys.\ Rev.\ D {\bf 98}, no. 8, 084041 (2018).
	
	\bibitem{Mazumdar:2018xjz} 
	A.~Mazumdar and G.~Stettinger,
	arXiv:1811.00885 [hep-th].
	
	
	\bibitem{Bravinsky}		
	A. O. Barvinsky and G. A. Vilkovisky, 
	Phys. Rept. 119 (1985) 1-74.
	
	\bibitem{Deser:2007jk} 
	S.~Deser and R.~P.~Woodard,
	Phys.\ Rev.\ Lett.\  {\bf 99}, 111301 (2007).
	
	\bibitem{Belgacem:2017cqo} 
	E.~Belgacem, Y.~Dirian, S.~Foffa and M.~Maggiore,
	JCAP {\bf 1803}, no. 03, 002 (2018).
	
	\bibitem{Tan:2018bfp} 
	L.~Tan and R.~P.~Woodard,
	JCAP {\bf 1805}, no. 05, 037 (2018).
	
	\bibitem{Deser:2013uya} 
	S.~Deser and R.~P.~Woodard,
	JCAP {\bf 1311}, 036 (2013).
	
	\bibitem{Conroy:2014eja} 
	A.~Conroy, T.~Koivisto, A.~Mazumdar and A.~Teimouri,
	Class.\ Quant.\ Grav.\  {\bf 32}, no. 1, 015024 (2015), [arXiv:1406.4998 [hep-th]].
	
	\bibitem{Woodard:2018gfj} 
	R.~P.~Woodard,
	Universe {\bf 4}, no. 8, 88 (2018).
	
	\bibitem{Moste} 
	V.~M.~Mostepanenko,
	J.\ Phys.\ Conf.\ Ser.\  {\bf 161}, 012003 (2009).
	
\end{thebibliography}
\end{document}